\documentclass[twocolumn,floatfix,amsmath,prl,aps,showpacs]{revtex4}

\usepackage{graphicx}
\usepackage{dcolumn}
\usepackage{bm}
\usepackage{psfrag}

\begin{document}


\title{Local Density of States for Individual Energy Levels in Finite Quantum Wires}

\author{Imke Schneider, Alexander Struck, Michael Bortz,  and Sebastian Eggert}

\affiliation{Department of Physics and Research Center OPTIMAS, 
Univ.~Kaiserslautern, D-67663 Kaiserslautern, Germany
}

\date{\today}

\begin{abstract}

The local density of states (LDOS) 
in finite quantum wires is calculated as a function
of discrete energies and position along the wire.
By using a combination of numerical density matrix
renormalization group (DMRG) calculations and analytical bosonization techniques
it is possible to obtain a good understanding of the 
local spectral weights along the wire in terms of the underlying 
many-body excitations.

\end{abstract}                                                                 

\maketitle

There has been increasing interest in quasi one-dimensional quantum wires 
over the last few decades.
Tunneling experiments 
have revealed signatures of strong correlations in different setups ranging from 
a powerlaw behavior of the temperature dependent 
conductance in carbon nanotubes \cite{Bockrath1999,Dekker1999} to being able to 
map out the momentum resolved 
tunneling in parallel quantum wires in a GaAs/AlGaAs
heterostructure \cite{Auslaender2005,Auslaender2002,Steinberg2006}.
By now the local tunneling density near nanotube edges can be measured fully resolved in 
energy and position \cite{Lee2004}, which has been interpreted as the interference pattern
of collective excitations.  
Standing waves at discrete energies
corresponding to the few lowest lying levels have also been observed
in finite tubes with screened interactions \cite{venema,lemay}.
It can be expected that similar experiments 
will be able to identify collective excitations 
in short finite wires if the screening from the substrate can be reduced.

The central quantity of interest is the 
local density of states (LDOS) of inserting one particle at 
position $x$ to reach an excited state $\langle \omega_n|$ 
with $N_0+1$ particles from the ground state $|N_0\rangle$
\begin{eqnarray}\label{ldos}
\rho(\omega,x)& =& \sum_{n} |\langle \omega_n| \psi^\dagger_x|N_0\rangle|^2\,
\delta(\omega-\omega_n) \nonumber \\
& =&  -\frac{1}{\pi} {\rm Im} \int_0^\infty e^{i\omega t} G^R(t,x)dt.
\end{eqnarray}
Using the Luttinger liquid formalism
it is now well understood how the scaling laws of the LDOS and the correlation functions 
are affected by boundaries \cite{Eggert1992,Fabrizio1995,Eggert1996,Mattsson1997,Meden2000,Eggert2000,paata}. 
The position resolved LDOS 
in finite wires was considered in Ref.~[\onlinecite{Anfuso2003}]
for the lowest few levels assuming a perfect degeneracy in the Luttinger liquid model.

From the numerical side it is surprising that so far 
only very few works \cite{Jeckelmann2004} 
were able to analyze the DOS 
in interacting lattice models exactly. 
Part of the difficulty appears to be that the 
LDOS in Eq.~(\ref{ldos}) 
is related to the time-dependent retarded Green's function $G^R(t)$. 
In dynamical DMRG 
methods dramatic progress has been made over the past few years. 
Recently, it was possible to determine 
the spectral function as a function of the photoemission wavevector \cite{Jeckelmann2004}.  
Unfortunately, individual levels have not yet been resolved due to a
finite correlation time or a finite cutoff in the 
Lehmann representation of the Green's function.
To our knowledge there have been no 
numerical calculations for the LDOS in interacting lattice models so far.

In this paper, we are now able to quantitatively calculate the LDOS of a finite 
quantum wire by using a combination of bosonization and DMRG techniques \cite{White1992}.  
In the numerical DMRG calculations we follow a direct approach, 
by targeting several excited states $\langle \omega_n|$
in the sector with one additional fermion on top 
of the ground state.
Keeping track of the particle number
and all the anticommuting 
transition operators $\psi^\dagger_x$
it is then possible to evaluate the matrix elements of the LDOS in Eq.~(\ref{ldos})
with very good spatial
and energy resolution compared to calculating the DOS via the Green's function.
Using bosonization 
a general analytic formula for the LDOS of nearly degenerate states is derived.
The combination of numerical and analytical calculations gives a good 
understanding of the distribution of spectral weights over individual states.
The well-known powerlaws, however,  can only be observed after 
a summation over nearly degenerate states.
The effect of spin is also discussed.

\begin{figure}
   \begin{center}
        \includegraphics[width=0.47\textwidth]{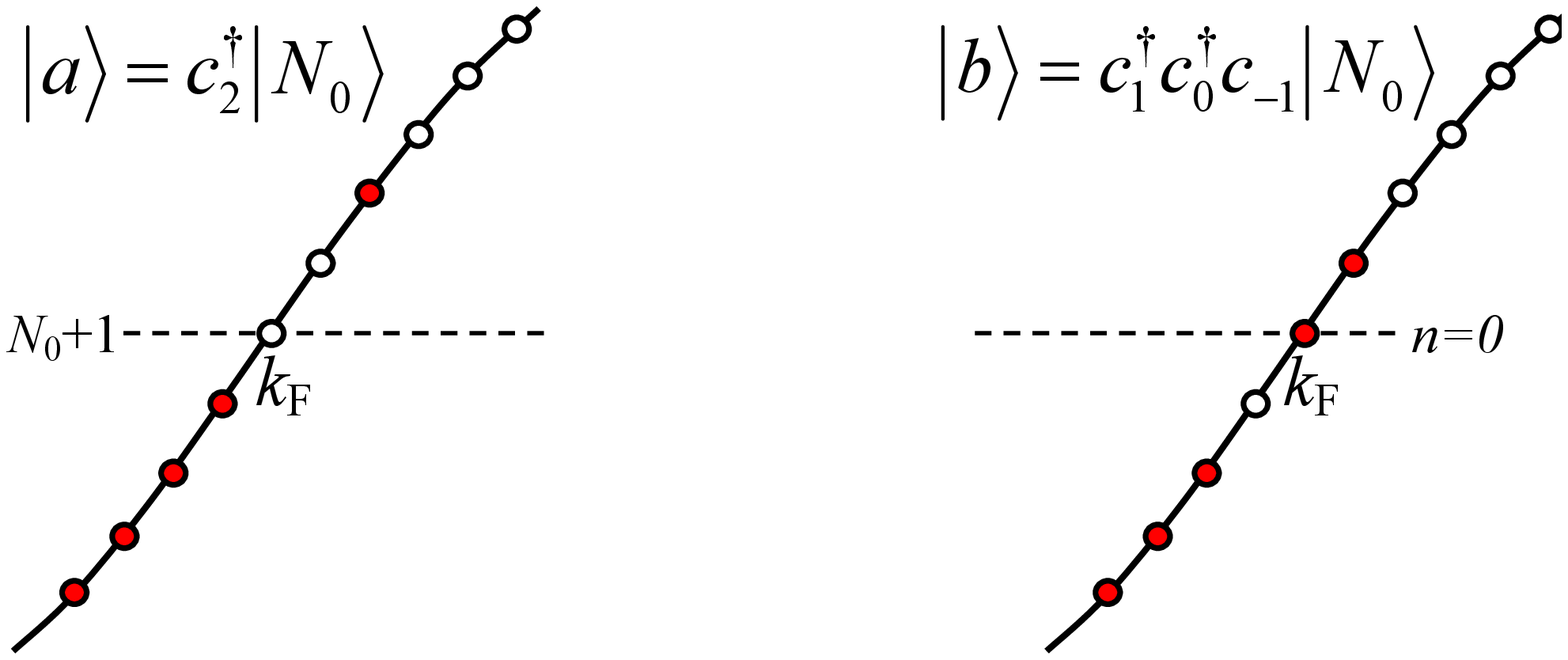}
   \end{center}
    \caption{(color online) Single particle excitations on the 
filled Fermi sea for the non-interacting system.} 
\label{single_p_e}
\end{figure}

 We consider 
interacting spinless fermions hopping on a finite wire
with $L$ sites and ''open'' ($\psi_0=\psi_{L+1}=0$) boundary conditions 
\begin{equation}
H= -t \sum_{x=1}^{L-1}\left(\psi^\dagger_x \psi^{\phantom \dagger}_{x+1} + h.c.\right)
+ U \sum_{x=1}^{L-1}
n_x n_{x+1},
\label{ham}
\end{equation}
where $n_x = : \!\!\psi^\dagger_x \psi^{\phantom \dagger}_{x}\!\!:$.
This is the simplest model that illustrates Luttinger liquid physics
for $-2t<U \leq 2t$ and is often used to describe quantum wires. 
It is believed that the results can be generalized to systems with spin
as will be discussed later.
Tunneling one single particle (or hole) into the wire will create excited states
that may in general involve several more fermions (i.e.~a so-called "single particle"
excitation may still be a many-body entangled state).  
In order to illustrate the nature of such many-body excitations, consider
the system without interaction $(U=0)$ first.  In that case, eigenstates are created 
from the vacuum by  creation operators
$c_n^\dagger = \sqrt{\frac{2}{L+1}}\sum_x \psi_x^\dagger \sin (k_n+k_F) x $,
where the wavevector $k_n=  n \frac{\pi}{L+1}$ 
is measured relative to the Fermi point $k_F$.
For convenience 
$k_F=(N_0+1)\frac{\pi}{L+1}$
is chosen so that the number of particles 
in the filled Fermi sea is given by $N_0$ with energy $E_0$
and there is no charging energy to 
the $N_0+1$ particle sector as indicated in Fig.~\ref{single_p_e}.
A typical example of a single-particle excitation in a
tunneling experiment is the creation of one fermion 
above the filled Fermi sea $|a\rangle=c_2^\dagger|N_0\rangle$
as shown in Fig.~\ref{single_p_e}. 
In that case, the LDOS in Eq.~(\ref{ldos}) is simply given by the 
square of the corresponding standing wave.
A second example of a possible excitation 
 $|b\rangle =c_1^\dagger c_0^\dagger c_{-1}|N_0\rangle$ is 
shown in Fig.~\ref{single_p_e}, where also one additional fermion is created
and another one has been 
excited from the Fermi sea.  
For non-interacting fermions 
the overlap matrix elements $\langle b|\psi^\dagger_x|N_0\rangle$ 
of this many-body excited state vanish 
in the LDOS Eq.~(\ref{ldos}),
but this is no longer true for the interacting case.
Using numerical results and bosonization we are now able to 
describe quantitatively how spectral weight is shifted from ''simple  
excitations''  (type $|a\rangle$) to many-body excited states (type $|b\rangle$)
due to interactions and how this is reflected in the standing waves of the LDOS.

Note, that  if the dispersion relation was exactly linear,
the states $|a\rangle$ and $|b\rangle$ would be degenerate. 
In fact a linear spectrum is often assumed, because in that case
all many-body excitations are exactly quantized with energy levels 
$\omega_n=v_F k_n = \frac{v_F \pi}{L+1}n$ relative to $E_0$, where 
the number of possible partitions of $n = \sum_{\ell=1}^{\infty}\ell m_\ell$ into 
integers $m_\ell$ gives the degeneracy of the level $\omega_n$
(e.g. for $n=3=2+1=1+1+1$ there
are three possible partitions).
However, numerically  we keep track of the exact spectrum of the
finite lattice model, which is never exactly linear so that the degeneracies are
lifted. 

Even without assuming a linear spectrum the problem can be bosonized 
by defining shifting operators
 $\varrho_n =\sum_{\ell}c_{\ell+n}^{\dagger} c_{\ell}^{\phantom\dagger}$
for $n \neq 0$,
which obey bosonic commutation relations
$\left[\varrho_{-n},\varrho_{n^\prime}\right] = n \delta_{n,n^\prime}$
on the infinitely continued spectrum \cite{egg07}. 
The zero mode $\varrho_0=\sum_\ell 
\left(c_{\ell}^{\dagger} c_{\ell}^{\phantom\dagger} - 
\langle c_{\ell}^{\dagger} c_{\ell}^{\phantom\dagger}\rangle_{k_F}\right)$ 
counts the number of particles relative to the filled Fermi sea.
It is possible to 
create any fermion state with a given particle number 
in terms of the $\varrho_n$.
For example, we find by combinatorial methods
that the addition of one fermion is represented by
the following superposition of many-body bosonic states
\begin{eqnarray}\label{cj}
c_n^\dagger|N_0\rangle=
\sum_{\sum_{\ell}\ell m_{\ell}=n} \prod_{\ell}
\frac{1}{m_\ell!} \left(\frac{\varrho_{\ell}}{\ell}\right)^{m_\ell}
 |0\rangle,
\end{eqnarray}
where $|0\rangle$ represents the ground state with $N_0+1$ fermions.
Here, the sum runs over occupation numbers 
$m_{\ell} = \langle \varrho_\ell \varrho_{-\ell}\rangle/\ell$
which correspond to all possible
partitions of $n =  \sum_{\ell}\ell m_{\ell}$ \cite{schoenhammer}. 
In particular, the states in Fig.~\ref{single_p_e} are given by 
\begin{eqnarray}\label{bs}
|a/b\rangle & = &  
\tfrac{1}{2}\left(\varrho_1^2\pm\varrho_2\right) |0\rangle.
\end{eqnarray}

The local addition of one fermion is described by  $\psi^\dagger_x \approx 
e^{-ik_F x} \psi^\dagger_R(x) - e^{ik_F x} \psi^{\dagger}_R(-x)$, where
\begin{eqnarray}
\psi_R^\dagger(x)=
c(x)
\exp{(\sum_{n>0} {e^{-ik_n x}}\frac{\varrho_{n}}{n} )}
\exp{(\sum_{n<0} {e^{-ik_n x}}\frac{\varrho_{n}}{n} )}
\label{vertex}
\end{eqnarray}
in accordance with open boundary 
conditions \cite{Eggert1992,Fabrizio1995,Eggert1996,Mattsson1997}. 
Here, the prefactor $c(x)=\frac{i}{\sqrt{2(L+1)}} 
e^{i({\phi}_0-\varrho_0\frac{\pi x}{L+1})}$ also contains the zero mode operators 
$\varrho_0$ and $\phi_0$,
which exactly create the $(N_0+1)$-particle ground state
$|0\rangle =e^{i({\phi}_0-\varrho_0\frac{\pi x}{L+1})}|N_0\rangle$.
The ambitious reader may enjoy verifying that the state in Eq.~(\ref{cj}) is normalized 
and that the free fermion result $|\langle N_0|c_n \psi^\dagger_x|N_0\rangle|^2
=\frac{2}{L+1}| \sin (k_F+k_n) x|^2$ can also be evaluated in the boson language
using Eq.~(\ref{cj}) and (\ref{vertex}).  

Interactions $U$ can now be treated by a Bogoliubov transformation 
${\varrho}_n = \alpha {\tilde \varrho}_n + \beta {\tilde \varrho}_{-n}$
where the rescaling parameters $\alpha=\frac{1}{2}(1/\sqrt{K}+\sqrt{K})$ 
and $\beta=\frac{1}{2}(1/\sqrt{K}-\sqrt{K})$
are characterized by the Luttinger liquid parameter $K$.  
For the model in Eq.~(\ref{ham}), $K$ and the Fermi velocity $v_F$ are known exactly 
as a function of $U$ by comparison with Bethe ansatz,
e.g.~at half filling $1/2K=1-\arccos(U/2t)/\pi$ and $v_F = 2t K \sin(\pi/2K)/(2K-1)$.
The interacting Hamiltonian becomes diagonal
$H=\frac{\pi {v}_F}{L+1}\sum_{\ell=1}^{\infty}\ell 
 {\tilde \varrho}_\ell {\tilde \varrho}_{-\ell}$ up to  non-linear
corrections, which are described by higher order operators.
After normal ordering the  prefactor of the transformed
fermion operator in Eq.~(\ref{vertex})    is given by
\begin{equation}
c(x) = \tfrac{i}{\sqrt{2(L+1)}} \left(\tfrac{\pi}{L+1}a\right)^{\beta^2}
\left(2 \sin \tfrac{\pi }{L+1}x\right)^{\alpha \beta} ,
\label{prefactor}
\end{equation}
where the inconsequential zero modes are left out.
Here $a$ is a so-called cutoff parameter of order one lattice constant,
which comes from a sum over physical 
states and can be determined numerically. 
\begin{figure}
   \begin{center}
        \includegraphics[width=0.47\textwidth,angle=0]{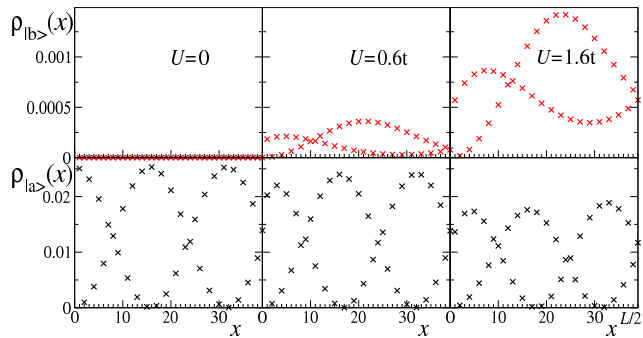}
   \end{center}
    \caption{(Color online) DMRG data for the LDOS 
$|\langle b| \psi^\dagger(x)|N_0\rangle|^2$ (first row) and 
$|\langle a| \psi^\dagger(x)|N_0\rangle|^2$ for $L=78$,
where $|b\rangle$ and $|a\rangle$ are eigenstates that have evolved from
the $n=2$ states 
in Fig.~\ref{single_p_e} when interactions $U$ are switched on. 
The spectral weight is shifted towards $|b\rangle$ with increasing interactions,
but $|a\rangle$ remains dominant.}
   \label{wavefunctions}
\end{figure}

The transformed boson vacuum and all eigenstates correspond to  complicated 
many-fermion superpositions, which we can now analyze numerically.  
We implement the model in Eq.~(\ref{ham}) with
 $L=78$ sites and $N_0+1 = 40$ using a multitarget DMRG. By keeping 
reflection and particle number symmetries, the position resolved 
matrix elements $\langle E |\psi^\dagger_x |N_0\rangle$ 
of 67 excited states $\langle E|$ are calculated.
The eigenenergies $E$ agree to at least 8 digits with the exact Bethe ansatz.

We use the simplest example of the states $|a\rangle$ and $|b\rangle$
in order to illustrate how the LDOS changes with $U$ as shown in 
Fig.~\ref{wavefunctions}.
The spectral weight of the many-body state $|b\rangle$ 
increases with interactions.  The wavefunction can be analyzed 
with the help of Eq.~(\ref{vertex}), so that the eigenstates are
known in terms of bosonic excitations,
e.g.~$|b\rangle\approx 
(0.541 \tilde\varrho_1^2-0.455 \tilde\varrho_2)|0\rangle$ for $U=1.6t$.

\begin{figure}
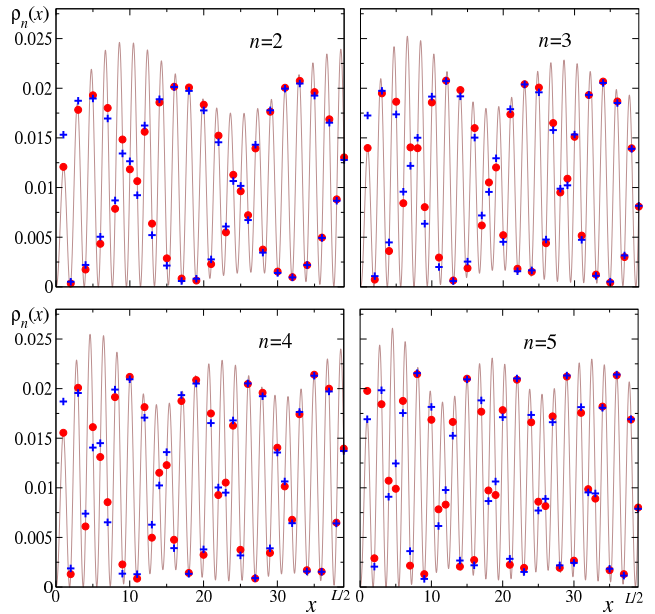

   \begin{center}
        \includegraphics[width=0.47\textwidth]{ldos23.eps}
        \includegraphics[width=0.47\textwidth]{ldos45.eps}
   \end{center}
\caption{(Color online)  DMRG data for the total LDOS of different levels $n$ 
for $U=1.4t$  
(blue crosses) compared to the bosonization results 
in Eq.~(\ref{ldos_analytical})  with $a=0.56$ (solid line). 
Red dots mark the LDOS from bosonization at discrete lattice points.  }
   \label{ldos0.7}
\end{figure}

The exact bosonic states are determined by higher order operators 
in the bosonized Hamiltonian, which are not known quantitatively.
However, since the complete nearly degenerate subspace 
for a given energy  $\omega_n=v_F k_n$ is known,
it is in principle possible to 
calculate the sum of the LDOS of nearly degenerate states, i.e. 
$\rho_n(x) =\int_{\omega\approx \omega_n}\rho(\omega,x) d\omega$. 
By summing over
bosonic eigenstates  $\prod_{\ell=1}^n
 \left(\ell^{m_\ell} m_\ell!\right)^{-1/2} \tilde\varrho_\ell^{m_{\ell}} |0\rangle$,
corresponding to all partitions of $n =  \sum_{\ell}\ell m_{\ell}$, we have found a
general analytic expression for this ''total LDOS''
 \begin{eqnarray}\label{ldos_analytical}
\rho_n(x)=|c(x)|^2 
{\sum_{n =  \sum\ell m_{\ell}}}
& &\left| \prod_{\ell=1}^n
\frac{\chi_{\ell}^{m_{\ell}}}{\sqrt{{\ell} m_{\ell}!}}
e^{ik_Fx}-{c.c.}\right|^2
\end{eqnarray}
where  $\chi_{\ell}(x)= \alpha e^{i k_{\ell} x}+ \beta e^{-ik_{\ell} x}$.
This formula is the analytic  generalization of the LDOS that previously was only known
for the first three levels \cite{Anfuso2003}. The total LDOS shows a pattern of 
standing waves that are modified by collective bosonic excitations 
in the form of characteristic density modulations as shown in 
Fig.~\ref{ldos0.7}.  
The analytical result in Eq.~(\ref{ldos_analytical}) can be compared with the
numerical data as shown in Fig.~\ref{ldos0.7} for the example of levels
up to  $n=5$  for 
$U=1.4t$ ($K\approx 0.67$).  The overall scale is the only adjustable parameter, which 
determines the cutoff in Eq.~(\ref{prefactor}), that ranges from
$a=  0.3\, ...\, 0.7$ as $U=0.2\, ... \, 1.8t$ is increased.
Note, that the lattice structure of the model (\ref{ham}) 
is incommensurate with the rapid oscillations of the LDOS, which 
leads to a beating of the signal.
The quantitative agreement with the analytical prediction 
is very good up to small deviations near the boundary. 
\begin{figure}
   \begin{center}
        \includegraphics[width=0.47\textwidth,angle=0]{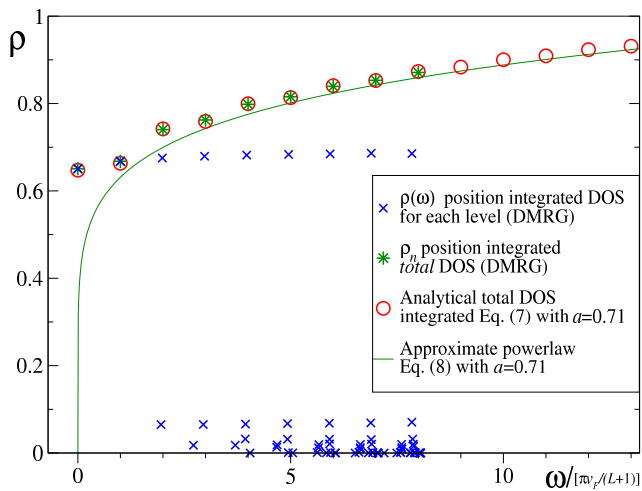}
   \end{center}
    \caption{(Color online) DMRG data for $U=1.8t$ and $L=78$
of the position integrated DOS 
$\rho(\omega)=\sum_{x}\rho(\omega,x)$ for each individual 
state (blue crosses).  The green stars represent the sum over 
all nearly degenerate states at each discrete energy $\omega_n$, which 
agrees very well with the integral of Eq.~(\ref{ldos_analytical}) (circles).}
\label{dos}
\end{figure}

The number of nearly degenerate states at each energy $\omega_n$ grows
rapidly with the number of partitions $\sim \exp(\pi\sqrt{2 n/3})/4 n \sqrt{3}$
and each of those states has an increasingly complex wavefunction, so that
a description in terms of approximate powerlaws is normally used.
Indeed, if the position dependence is integrated over, the total DOS
$\rho_n=\sum_x \rho_n(x)$ in Eq.~(\ref{ldos_analytical}) should
follow an approximate powerlaw
\begin{equation}
\rho_n \approx  \tfrac{1}{\Gamma(1+2 \beta^2)}\left(\tfrac{\pi a}{L+1}n\right)^{2 \beta^2} 
= \tfrac{1}{\Gamma(1+2 \beta^2)}
\left(\tfrac{a}{v_F} \omega_n\right)^{2\beta^2}.
\end{equation}
However, this only holds for the sum over all nearly degenerate states.  The 
individual states on the other hand show a much more interesting energy dependence
of the position integrated DOS
as shown in Fig.~\ref{dos}.  The dominant spectral weight in
each multiplet has evolved from the original simple 
excitation in Eq.~(\ref{cj}) (type $|a\rangle$), which has a surprisingly flat 
energy dependence.  The many-fermion states (type $|b\rangle$) 
are much more numerous, but have a much smaller DOS. 

The position integrated {\it total} DOS of each level in Fig.~\ref{dos} agrees well 
with the integral of Eq.~(\ref{ldos_analytical}).  There is no sign of any deviations
from the low energy theory, which should break down as the cutoff is 
approached $\omega \sim v_F/a$, i.e.~before the DOS reaches the non-interacting 
value of unity.  The alternation of the analytical results relative to the powerlaw
is due to boundary effects, which induce additional oscillations with
$\sin 2 \omega x/v_F$ \cite{Eggert2000}.
In general, for longer range interactions it is also possible to 
use a momentum dependent Luttinger liquid parameter \cite{longrange}.

The situation as shown in Fig.~\ref{dos} is actually quite generic for interacting
systems also in higher dimensions:  The single particle spectral weight is 
spread over many-body excitations, thereby renormalizing position, width, and 
total weight.  Our results show quantitatively how an effective
 finite lifetime of quasiparticle states evolves due to a spread over many-body states.

In one-dimensional systems with spin degrees of 
freedom $\sigma=\uparrow,\downarrow$,
 the excitations can be described
by two independent bosonic theories for spin and charge \cite{egg07}.
The spectrum is again given in the form of nearly degenerate multiplets
$\omega=\omega_{n_s}+\omega_{n_c}$, but now
labeled by two quantum numbers $(n_s,\ n_c)$
for spin and charge with
$\omega_{n_s}=v_{s} k_{n_s}$ and $\omega_{n_c} = v_{c} k_{n_c}$, where
$v_c=v_s$ in the non-interacting case.
A single particle state $c_{n,\sigma}^\dagger|N_0\rangle$ 
is now divided over all 
spin and charge multiplets $(n_s,\ n_c)$
corresponding to the different possibilities to
write the sum $n={n_s}+{n_c}$.  
However, for a given $n={n_s}+{n_c}$ the 
multiplets $(n_s,\ n_c)$ may now be of 
quite different energy, since
in general $v_c > v_s$ for repulsive interactions.
The spectral weight is therefore spread into
several {\it non-degenerate} spin and charge multiplets,
which is fundamentally different
from higher dimensional systems.
Moreover, the degeneracy within each multiplet is also lifted in the same
fashion as above.
In each multiplet there is again exactly one dominant state. 
The LDOS also shows a very interesting superposition of 
spin and charge density waves \cite{Anfuso2003}, which can be
calculated in detail with rescaled waves $\chi_\ell(x)/\sqrt{2}$ in 
Eq.~(\ref{ldos_analytical}).
However, the
smoking gun for Luttinger liquid behavior in a finite wire would 
be to identify the increasing number of states
due to spin-charge separation and the degeneracy splitting
with increasing $\omega$, 
which is very different from an equally spaced
single particle spectrum.
Numerically, however, the important effects of interactions can be much better illustrated
using the single channel (spinless) case discussed here, which avoids the complications of 
a huge number of states and a superposition of spin and charge standing waves.

In summary, we were able to describe the LDOS for finite quantum wires 
in detail by a combination of analytical and numerical methods.  
The analytical results allow the calculation of the total LDOS in each multiplet.
The DMRG calculations give the detailed distribution of local spectral weights 
over all many-body states in the low energy region.  
The combination of both methods shows explicitly 
how the wavefunctions and boson representations of
excited states evolve as interactions are turned on.
The cutoff parameter $a(U)$ and the normalization in Eq.~(\ref{prefactor})
can be determined from DMRG for the standard model in Eq.~(\ref{ham}).
Our results show that powerlaws are not sufficient to adequately describe the low
energy behavior of individual levels.  Instead, a large number of discrete states
with varying spectral weights and oscillating wavefunctions would be the 
generic signature of Luttinger liquid behavior in finite wires.  
The number of many-body
states with non-zero spectral weight is small for the first few levels, but 
increases exponentially with energy.  

\textbf{Acknowledgments}
We are grateful to M.~Andres, S.~Reyes and S.~S\"offing for helpful discussions and 
the SFB/Transregio 49 for financial support. I.~S.~acknowledges support 
of the Graduate Class of Excellence MATCOR funded by the State of Rheinland-Pfalz, Germany.

\end{document}